\documentclass[twocolumn,showpacs,prl]{revtex4}
\usepackage{amssymb}
\usepackage[intlimits]{amsmath}
\usepackage[dvipdfm]{graphicx}
\usepackage{dcolumn}
\usepackage{amsfonts}
\usepackage{bm}

\begin{document}

\title{Asymmetry dependence of proton correlations.}
\author{R. J. Charity$^{1}$, L. G. Sobotka$^{1,2}$, W. H. Dickhoff$^{2}$}
\affiliation{Departments of Chemistry$^{1}$ and Physics$^{2}$, Washington University, St.
Louis, Missouri 63130.}
\date{\today}

\begin{abstract}
A dispersive optical model analysis of \textit{p}+$^{40}$Ca and \textit{p}+$%
^{48}$Ca interactions has been carried out. The real and imaginary
potentials have been constrained from fits to elastic scattering data,
reaction cross sections, and level properties of valence hole states deduced
from $(e,e^{\prime }p)$ data. The surface imaginary potential was found to
be larger overall and the gap in this potential on either side of the Fermi
energy was found to be smaller for the neutron-rich p+$^{48}$Ca system.
These results imply that protons with energies near the Fermi surface
experience larger correlations with increasing asymmetry.
\end{abstract}

\maketitle

In the independent-particle model, nucleons in the nucleus move in a
mean-field potential generated by the other nucleons. All nucleon levels up
until the Fermi energy ($E_{F}$) are fully occupied, while those above are
empty. Although this model enables an understanding of various aspects of
nuclear structure, a full description of nuclei and nuclear matter requires
consideration of the correlations between the nucleons. These include
short-range, central and tensor interactions and longer range correlations
associated with low-lying collective excitations \cite{Dickhoff04}. As a
result, for closed-shell nuclei, single-particle (sp) levels below $E_{F}$
have an occupancy of only 70-80\% and the levels at higher energy have a
nonzero occupancy \cite{Pandharipande97}. The strength of the sp levels are
spread over energy, with narrow peaks or broad distributions (depending on
their separation from $E_{F}$). In addition, there is strength at very high
momentum \cite{Rohe04}.

Although there are numerous studies of the effect of correlations on the
properties of sp levels for nuclei near stability, there are only a few
studies for very neutron or proton-rich nuclei. From neutron knock-out
reactions, Gade \textit{et al.} \cite{Gade04} infer the occupancy of the $%
0d_{5/2}$ neutron hole state in the proton-rich $^{32}$Ar nucleus is
considerably reduced relative to those for stable nuclei.

An alternate method to study sp strength is through the use of the
dispersive optical model (DOM) developed by Mahaux and Sartor \cite{Mahaux91}%
. This description employs the Kramers-Kronig dispersion relation that links
the imaginary and real parts of the nucleon self-energy \cite{Dickhoff05}.
This procedure links optical-model (OM) analyses of reaction data at
positive energies to structural information at negative energies. In the
present work, the properties of proton levels in Ca nuclei as a function of
asymmetry $\delta $=$\frac{N-Z}{A}$ are investigated with the DOM.
Previously measured elastic-scattering and reaction-cross-section data for
protons on $^{40}$Ca and $^{48}$Ca as well as level properties of hole
states in these nuclei, inferred from ($e,e^{\prime }p$) reactions, were
simultaneously fit. The dependence on $\delta $ is extracted and used to
predict level properties of $^{60}$Ca.

\begin{figure}[tbp]
\includegraphics*[scale=0.55]{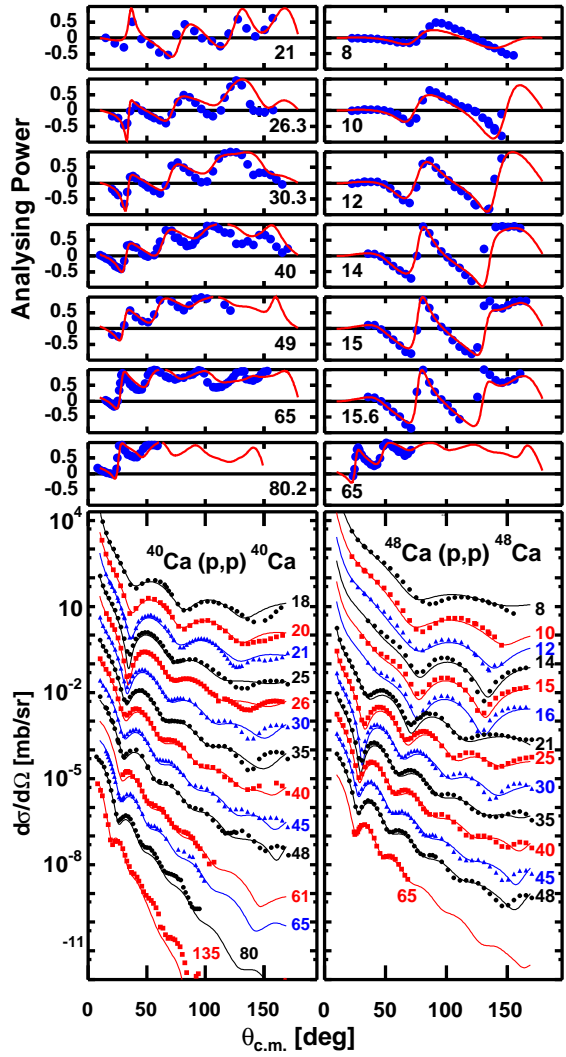}
\caption{(Color online) Calculated and experimental elastic-scattering
angular distributions of the differential cross section $d\protect\sigma %
/d\Omega $ and analyzing power. Left panels shows results for \textit{p}+$%
^{40}$Ca, right for \textit{p}+ $^{48}$Ca. Each curve is labeled by the
proton energy in MeV. For display purposes, successively higher energy
curves and data for $d\protect\sigma /d\Omega $ are scaled down by an
additional factor of 4.5. For the lowest energies, compound-elastic
contributions were also included in the fits. }
\label{fig:elast}
\end{figure}

In the DOM, the complex energy-dependent potential felt by the protons is
comprised of a real $\mathcal{V}$, volume $W_{v}$ and surface $W_{s}$
imaginary components, plus spin-orbit $V_{so}$ and Coulomb $V_{c}$
potentials,%
\begin{multline}
\mathcal{U}(r,E)=-\mathcal{V}(r,E)+V_{so}(r)+V_{c}(r) \\
-iW_{v}(E)\;f(r,r_{v},a_{v})+4ia_{s}W_{s}(E)\frac{d}{dr}f(r,r_{s},a_{s}).
\end{multline}%
Wood-Saxon form factors $f(r,r_{i},a_{i})=[1+e^{\frac{r-r_{i}A^{1/3}}{a_{i}}%
}]^{-1}$ are used. The real part of the nuclear potential is assumed to be
given by two terms 
\begin{equation}
\Delta \mathcal{V}(r,E)=V_{HF}(E)\;f(r,r_{HF},a_{HF})+\Delta \mathcal{V}(r,E)
\end{equation}%
where $V_{HF}$ has a smooth energy dependence arising from the nonlocality
or momentum-dependence of the microscopic self-energy. The dispersive
correction $\Delta \mathcal{V}$ has volume and surface parts, 
\begin{equation}
\Delta \mathcal{V}(r,E)=\Delta V_{v}(E)f(r,r_{v},a_{v})+4a_{s}\Delta V_{s}(E)%
\frac{d}{dr}f(r,r_{s},a_{s}),
\end{equation}%
and is related to the imaginary potential through the dispersion
relationship, i.e., 
\begin{equation}
\Delta V_{i}(E)=\frac{P}{\pi }\int_{-\infty }^{\infty }W_{i}(E^{\prime
})\left( \frac{1}{E^{\prime }-E}-\frac{1}{E^{\prime }-E_{F}}\right)
dE^{\prime }
\end{equation}%
where $i=v,s$ and $P$ stands for the principal value. The dispersive
corrections are a result of coupling to non-elastic channels. The surface
term accounts for the influence of low-lying collective states and giant
resonances.

The form for the imaginary potential must take into account the dominance of
surface and volume absorption at low and high positive energies,
respectively. In addition around $E_{F}$ the imaginary potentials must be
zero. The potential should be approximately symmetric around $E_{F}$, but
further away it must become asymmetric as there are a finite number of hole
states. In this work we assume 
\begin{equation}
W_{v}(E)=A_{v}\frac{(E-E_{F})^{4}}{(E-E_{F})^{4}+B_{v}^{4}}+\Delta W_{NM.}(E)
\end{equation}%
where the energy-asymmetric correction $\Delta W_{NM.}(E)$ is derived from
nuclear-matter considerations \cite{Mahaux91}. The surface potential was
taken as the difference of two functions that cancel at large energies, i.e.,%
\begin{gather}
\Delta W_{s}\left( E\right) =\omega (E,A_{s}^{1},B_{s}^{1},c,0)-\omega
(E,A_{s}^{2},B_{s}^{2},c,Q), \\
\omega (E,A,B,c,Q)=A\;\Theta \left( X\right) \frac{X^{4}}{X^{4}+B^{4}}%
e^{-cX},
\end{gather}%
where $\Theta \left( X\right) $ is Heavyside's step function, $X$ =$%
\left\vert E-E_{F}\right\vert -Q$, $A_{s}^{2}=A_{s}^{1}e^{-c\,Q}$, and $%
Q=B_{s}^{1}+\Delta B$.

The Hartree-Fock potential is often assumed to decrease linearly or
exponentially with energy. We took the form%
\begin{equation}
V_{HF}(E)=\frac{2A_{HF}}{1+\exp (B_{HF}\left( E-E_{F}\right) /A_{HF})}
\end{equation}%
which is approximately linear around $E_{F}$ and becomes more exponential at
larger energies. This form provided a reasonable location for the $0s_{1/2}$
level in $^{40}$Ca.

The parameters of the DOM were fit for both $^{40}$Ca and $^{48}$Ca from a
large set of data covering both positive and negative proton energies. For $%
^{40}$Ca, 14 experimental elastic-scattering angular distributions for
energies from 18 to 135 MeV\ \cite%
{Fulmer69,Dicello71,Oers71,Nadasen81,Sakaguchi82,McCamis86} and seven data
sets for the analyzing power measured at energies from 21 to 80 MeV \cite%
{Sakaguchi82,Blumberg66,Graig66,Watson67,Hnizdo71,Schwandt82} were included.
For $^{48}$Ca the fitted data included 14 angular distributions and seven
sets of analyzing power at energies from 8 to 65 MeV \cite%
{Sakaguchi82,McCamis86,Liers71,Lombardi72}. Reaction cross sections for both
targets were taken from the tabulations of Bauhoff \cite{Bauhoff86}. For the 
$0d_{5/2}$, $1s_{1/2}$, and $0d_{3/2}$ proton holes states, the mean level
energies, r.m.s. radii, spectroscopic factors, and widths inferred from
measured ($e,e^{\prime }p$) cross sections \cite{Kramer89,Kramer01} were
also included. Lastly, the fits considered the mean energies of the $%
0f_{7/2} $ and $0f_{5/2}$ (for $^{48}$Ca only) particle levels \cite%
{Millener73}.

\begin{figure}[tbp]
\includegraphics*[scale=0.4]{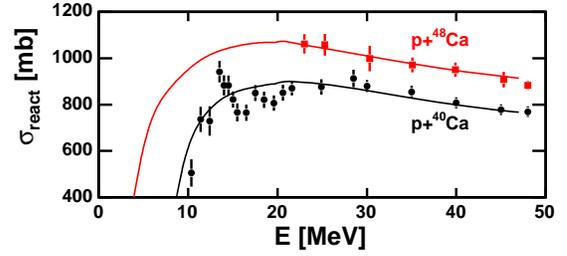}
\caption{(Color online) Total reaction cross sections are displayed as a
function of proton energy.}
\label{fig:react}
\end{figure}

\begin{figure}[tbp]
\includegraphics*[scale=0.4]{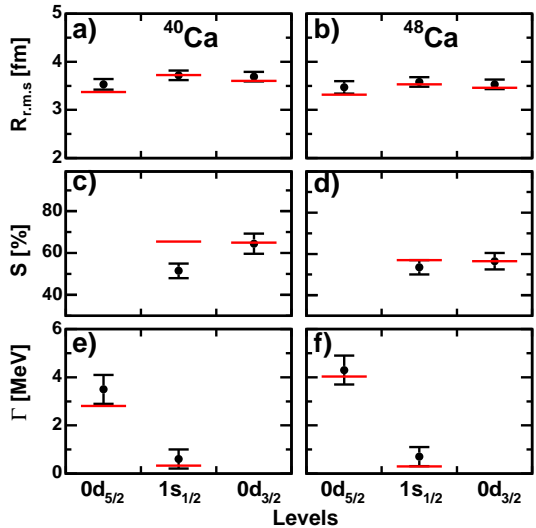}
\caption{(Color online) Fitted level properties for the $0d_{5/2}$, $%
1s_{1/2} $, and $0d_{3/2}$ proton hole states. The left (a,c,e) and right
(b,d,f) panels display the fits for $^{40}$Ca and $^{48}$Ca, respectively.
The fitted quantities include a,b) the root-mean-squared radius $R_{r.m.s}$,
c,d) the spectroscopic factor $S$ expressed as a percent of the
independent-particle-model value, and e,f) the widths $\Gamma $ of these
levels.}
\label{fig:level}
\end{figure}

The seven geometric parameters ($r_{HF},a_{HF}$=$%
a_{v},r_{v},r_{s},a_{s},r_{so},a_{so}$) defining the form factors were
varied, but kept identical for both targets. Similarly the fit parameters $%
B_{s}^{2}$, $\Delta B$ and $c$ defining the decay of the $W_{s}(E)$ at large
energies, as well the magnitude of the spin-orbit term, were also set
identical for both targets. Only $A_{HF}$, $B_{HF}$, $A_{s}^{1}$,$B_{s}^{1}$%
, $A_{v}$, and $B_{v}$ were allowed to differ for each isotope.

\begin{figure}[tbp]
\includegraphics*[scale=0.4]{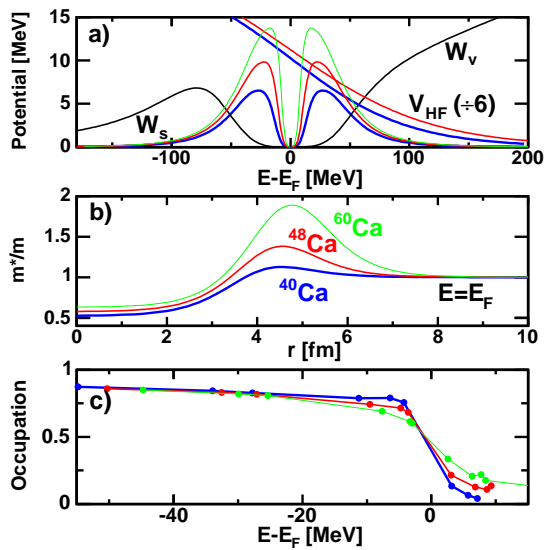}
\caption{(Color online) Quantities derived from DOM fits. a) The energy
dependence of the Hartree-Fock potential $V_{HF}$, and the surface $W_{s}$
and volume $W_{v}$ imaginary potentials. b) The radial dependence of the
effective mass at $E_{F}$. c) The occupation probabilities (points indicate
sp levels). For all plots, the thick (back), medium (red) and thin (green)
curves give the results for $^{40}$Ca, $^{48}$Ca, and $^{60}$Ca,
respectively.}
\label{fig:poten}
\end{figure}

The final fit to the experimental data is shown in Figs.~\ref{fig:elast}-\ref%
{fig:lev} and the fitted potentials are displayed in Fig.~\ref{fig:poten}a.
It was found possible to obtain similar quality fits with different
potentials, however a robust feature \textit{of all good fits} was that the
magnitude of the surface imaginary potential ( $A_{s}^{1})$ was larger and
its width $B_{s}^{1}$ (of the minimum around $E_{F}$) was narrower for the
neutron-rich $^{48}$Ca nucleus. There was however an ambiguity in
determining the rate at which this potential diminished at large energies.
This ambiguity is coupled with an ambiguity in determining the magnitude and
the rate of increase of $W_{v}\left( E\right) $.

Tornow \textit{et al.} \cite{Tornow90} fitted similar data for $^{40}$Ca
with the DOM using a $W_{s}\left( E\right) $ which decreased slowly and was
still substantial at the highest energy considered in the work. With such a
slow diminishing of $W_{s}\left( E\right) $, one can also obtain good fits
to the $^{48}$Ca data, however the fitted $W_{v}\left( E\right) $ potentials
are substantially different for $^{40}$Ca and $^{48}$Ca. On the other hand,
if $W_{s}\left( E\right) $ is made to diminish faster, these differences can
be reduced to zero.

Ambiguities in determining potentials in standard OM fits are well known,
however volume integrals of the potentials have been shown to be better
defined \cite{Mahaux91}. Comparisons of the imaginary volume integral, $%
J_{W}(E)=\int W(r,E)d\mathbf{r}$, obtained from the OM fits in the
referenced experimental studies, indicate that $J_{W}$ is larger below $%
E\sim $50~MeV in $^{48}$Ca. However for higher energies, there is no
discernible difference for the two isotopes. Thus for $^{48}$Ca, if its
larger surface potential is still significant in this higher energy region, $%
W_{v}\left( E\right) $ must be smaller to produce similar values of $J_{W}$.
We believe this result is artificial, and therefore a solution where $%
W_{s}\left( E\right) $ diminishes faster is preferable. In any case, it is
not possible to constrain any difference in $W_{v}\left( E\right) $ for the
two nuclei. Thus, for the final fit we have taken $W_{v}\left( E\right) $ to
be independent of asymmetry. This is consistent with global OM fits \cite%
{Becchetti69,Varner91} which have a significant asymmetry dependence for $%
W_{s}\left( E\right) $, but none for $W_{v}\left( E\right) $. Theoretically
some asymmetry dependence of $W_{v}\left( E\right) $ would be expected and
higher-energy $^{48}$Ca data would provide sensitivity to this.

\begin{figure}[tbp]
\includegraphics*[scale=0.4]{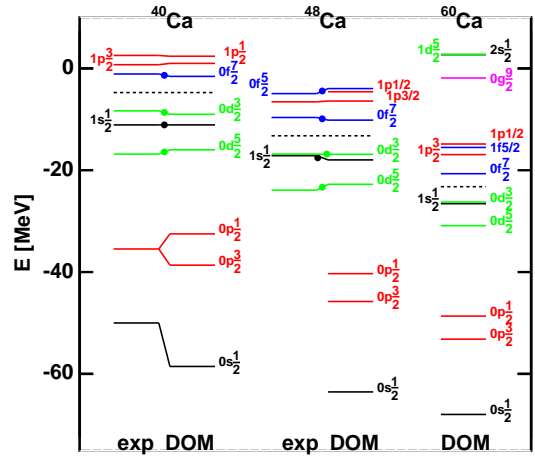}
\caption{(Color online) Comparison of experimental proton single-particle
levels \protect\cite%
{Kramer89,Kramer01,Millener73,James69,Youngblood70,Doll76,Watson89} to DOM
calculations. For the levels indicated with the solid dots, their energies
were included in the fits. The dashed lines indicate the Fermi energy.}
\label{fig:lev}
\end{figure}

Calculated and experimental sp level schemes for $^{40}$Ca and $^{48}$Ca are
displayed in Fig.~\ref{fig:lev}. Apart from the levels included in the fit
(indicated with the solid dots), the other known levels are well reproduced.
The $0s_{1/2}$ level of $^{40}$Ca is very wide and even though the DOM\
prediction for its energy is low, it lies within the experimentally
determined width \cite{James69}.

Present DOM calculations have been extrapolated to $^{60}$Ca assuming the
parameters $A_{HF}$, $B_{HF}$, $A_{s}^{1}$, and $B_{s}^{1}$ vary linearly
with $\delta $. The extrapolated energy-dependence of $W_{s}(E)$ is shown in
Fig.~\ref{fig:poten}a as the thin-lined curve and the predicted sp level
scheme is displayed in Fig.~\ref{fig:lev}. The surface dispersive correction
is large for this nucleus and its effects are quite apparent. The levels in
the immediate vicinity of $E_{F}$ are focused closer to $E_{F}$, increasing
the density of sp levels. A reduced gap between the particle and hole
valence levels implies that the closed-shell nature of this nucleus has
diminished and proton pairing may be important.\ The levels further from $%
E_{F}$ have been pushed away and as a result there are big gaps between the $%
0p_{1/2}$ and the $0d_{5/2}$ and also between the $1p_{1/2}$ and $0g_{9/2}$
levels. 

The occupation probabilities as defined in Ref.~\cite{Mahaux91} are plotted
in Fig.~\ref{fig:poten}c. The results for $^{48}$Ca agree with the
theoretical work of Ref.~\cite{Rijsdijk92}. For proton hole states just
below $E_{F}$, the occupation probabilities have decreased for the more
neutron-rich $^{48}$Ca while the opposite is true for the particle states.
This trend is further accentuated in our extrapolation to $^{60}$Ca. On the
other hand, the more deeply-bound $0s_{1/2}$, $0p_{3/2}$, and $0p_{1/2}$
levels show very little sensitivity. Their occupancies are more sensitive to
the volume imaginary component whose asymmetry dependence was not
constrained.

Our observation that the occupancies of valence proton hole states are
reduced in neutron-rich $^{48}$Ca can be compared to the reduced occupancy
of the valence neutron hole state in the proton-rich $^{32}$Ar inferred by
Gade \textit{et al.} \cite{Gade04}. Thus a preponderance of one type of
particle, reduces the occupancies of valence hole states for the other type.
This indicates that correlations are stronger for these valence nucleons.
Recent calculations of asymmetric nuclear matter by Frick \textit{et al.} 
\cite{Frick05} also predict such a result as a purely volume effect. Its
origin is probably similar in both cases reflecting that \textit{p-n}
interactions are stronger than \textit{n-n} or \textit{p-p, }partly because
of the tensor force. Thus, protons in neutron-rich systems are more strongly
correlated as illustrated from the inferred spectroscopic factors of 65\%,
56\% and 50\% for the $0d_{3/2}$ proton level in $^{40}$Ca, $^{48}$Ca and $%
^{60}$Ca, respectively. Conversely for neutrons, the opposite is expected.

The nucleon effective mass $m^{\ast }(r,E)/m=1-dV(r,E)/dE$ ($m$ is the
nucleon mass) at $E$=$E_{F\text{ }}$inferred from this work is displayed in
Fig.~\ref{fig:poten}b. Only the surface contribution around $r$=4-5~fm has
been constrained and it increases significantly with asymmetry. This
suggests an asymmetry dependence of the surface component of the
level-density parameter.

In conclusion, the properties of proton single-particle states in the
vicinity of $E_{F}$ for $^{40}$Ca and $^{48}$Ca have been studied with a
comprehensive dispersive-optical-model analysis of elastic-scattering and
bound-level data. The analysis indicates that the imaginary surface
potential is $\sim $50\% larger and the minimum around the Fermi energy is
narrower for the neutron-rich $^{48}$Ca nucleus. This implies that, with
increasing asymmetry, the occupancies of proton levels vary more smoothly
across the Fermi surface, a consequence of increased correlations. The
present observations and those of Gade \textit{et al.} \cite{Gade04} can be
understood from the larger strength of \textit{p-n} relative to the \textit{%
p-p} and \textit{n-n} interactions. Hence, protons (neutrons) experience
larger (weaker) correlations in neutron-rich matter. The reversed is true
for proton-rich matter.

This work was supported by the U.S. Department of Energy, Division of
Nuclear Physics under grant DE-FG02-87ER-40316.


\end{document}